\documentclass[prl,preprint]{revtex4}
\usepackage{amsmath}
\usepackage{amssymb}

\input{tcilatex}

\begin{document}

\title{Pair-breaking and superconducting state recovery dynamics in MgB$_{2}$%
}
\author{J. Demsar,$^{1,2}$ R.D. Averitt,$^{1}$ A.J. Taylor,$^{1}$ V.V.
Kabanov,$^{2}$ W. N. Kang,$^{3}$ H. J. Kim,$^{3}$ E. M. Choi,$^{3}$\ S. I.
Lee$^{3}$}
\affiliation{$^{1}$Los Alamos National Lab, MST-10, Los Alamos, NM 87545\\
$^{2}$"J.Stefan" Institute, Jamova 39, Ljubljana, Slovenia\\
$^{3}$National Creative Research Initiative Center for Superconductivity,
Department of Physics, Pohang University of Science and Technology, Pohang
790-784, Korea}
\pacs{PACS number}

\begin{abstract}
We present studies of the photoexcited quasiparticle dynamics in MgB$_{2}$
where, using femtosecond optical techniques, Cooper pair breaking dynamics
(PBD) have been temporally resolved for the first time. The PBD are strongly
temperature and photoexcitation intensity dependent. Analysis of the PBD
using the Rothwarf-Taylor equations suggests that the anomalous PBD arises
from the fact that in MgB$_{2}$ photoexcitation is initially followed by
energy relaxation to high frequency phonons instead of, as commonly assumed,
e-e thermalization. Furthermore, the bare quasiparticle recombination rate
and the probability for pair-breaking by phonons have been determined.
\end{abstract}

\maketitle

In recent years, femtosecond real-time spectroscopy has been shown to
present an excellent experimental alternative for studying temperature (T)
dependent changes in the low energy electronic structure of strongly
correlated electron systems\cite%
{Kabanov,YBCOOver,THzAveritt,Segre,Schneider,CMR,bb,HFPaper}. In these
experiments, a femtosecond laser pump pulse excites electron-hole pairs via
an interband transition in the material. In a process which is similar in
many materials including metals and superconductors (SC), these hot carriers
rapidly thermalize via electron-electron ($e-e$) and electron-phonon ($e-ph$%
) collisions reaching states near the Fermi energy within 10-100 fs. The
subsequent relaxation and recombination dynamics (strongly affected by the
opening of the superconducting\cite%
{Kabanov,YBCOOver,THzAveritt,Segre,Schneider} or charge density wave\cite{bb}
gap), are observed either by measuring photoinduced (PI) changes in
reflectivity or transmission at optical frequencies\cite%
{Kabanov,YBCOOver,Segre,Schneider,bb}, or by measuring conductivity dynamics
at terahertz (THz) frequencies.\cite{THzAveritt} While high-temperature
superconductors have been extensively studied over the last decade or so\cite%
{Kabanov,YBCOOver,THzAveritt,Segre,Schneider}, the data on more conventional
BCS type superconductors is fairly limited\cite{Federici,Carr}. To our
knowledge, no systematic study of photoexcited quasiparticle (QP) dynamics
with femtosecond resolution has been performed on low-T$_{c}$
superconductors.

In this Letter, we present the first femtosecond time-resolved study of
photoexcited carrier dynamics in the recently discovered superconductor MgB$%
_{2}$, utilizing both optical pump $-$ THz probe (OPTP) and optical pump $-$
optical probe (OPOP) techniques. The aim of this study is to elucidate the
photoexcited carrier dynamics in MgB$_{2}$ and in SC's in general, to
compare the dynamics to those in cuprates, and to obtain new and
complementary knowledge of the electronic structure of MgB$_{2}$. The
discovery of bulk superconductivity below $39$ K in MgB$_{2}$\cite{Nagamatsu}
has generated a great deal of excitement since $T_{c}$ is higher by nearly a
factor of two in comparison to other previously known simple intermetallic
superconductors. While the observation of a significant boron isotope effect%
\cite{isotope}, and\ the spin-singlet nature of the pairing\cite{NMR}
indicate conventional phonon mediated BCS pairing in MgB$_{2}$, several
experimental\cite{Cp,STM,ARPES2} and theoretical studies\cite{Liu,Mazin,Choi}
suggest suggest the presence of two distinct energy gaps\cite{Cp}.

Since the SC gaps in MgB$_{2}$ lie in the THz range\cite{KaindlPRL}, OPTP
spectroscopy is well-suited to study the dynamics of PI quasiparticles. In
particular, the Cooper pair-breaking dynamics (PBD) have been time-resolved
for the first time. The PBD are strongly temperature and photoexcitation
fluence dependent. This is attributed to, following photoexcitation, an
initially strong relaxation to high frequency phonons, and supported by
detailed analysis in terms of a phenomenological Rothwarf-Taylor model\cite%
{RothwarfTaylor}. The SC state recovery dynamics, on the other hand, proceed
on the timescale of hundreds of picoseconds. Systematic studies as a
function of temperature, excitation intensity, and film thickness suggest
that pair recovery is governed by the anharmonic decay of 2$\Delta $
acoustic phonons.

In time-domain THz spectroscopy a nearly single-cycle electric field
transient, containing Fourier components from $\sim 100$ GHz to several THz
is generated via optical rectification of 150 fs optical pulses in a ZnTe
crystal. The THz electric field transmitted through a sample, E$_{sam}$(t),
is detected using the Pockels effect in ZnTe\cite{THzAveritt,CMR}. A
measurement of the transmitted electric field is also made using a suitable
reference, $E_{ref}(t)$, which, for these experiments, is a blank sapphire
substrate. Dividing the Fourier transforms of the time domain sample and
reference data gives the complex transmissivity $T(\omega )=E_{sam}(\omega
)/E_{ref}(\omega )$. The real and imaginary conductivities ($\sigma
_{r}(\omega )$ , $\sigma _{i}(\omega )$) of the film are determined using
the appropriate complex Fresnel equation, without the need for
Kramers-Kronig analysis \cite{Nuss}. For the OPTP experiments, the induced
change in the transmitted electric field, $\Delta E_{sam}(t)$, is measured
as a function of time-delay with respect to an optical excitation pulse.
From $\Delta E_{sam}(t)$, it is then possible to determine the PI change of $%
\sigma (\omega )$ with picosecond (ps) resolution\cite%
{CMR,Heinz,Schmuttenmaer}.

The OPTP experiments were performed on 80 nm and 100 nm MgB$_{2}$ thin films
($T_{c}\sim 34$ K) on sapphire\cite{Kang}, whereas the films used in OPOP%
\textbf{\ }experiments had thicknesses of 300 and 400 nm, and $T_{c}=39$ K ($%
\Delta T_{c}\sim 0.15K$). Further details of the film growth\cite{Kang} and
a detailed description of experimental techniques are given elsewhere\cite%
{Kabanov,YBCOOver,THzAveritt}. The photoexcitation fluence \emph{F} ranged
from $0.1-5$ $\mu $J/cm$^{2}$, corresponding\cite{IntJModPhys} to an
absorbed energy density $\Omega =2-110$ $\mu $eV/unit cell. For comparison, $%
\Omega $ corresponding to complete destruction of SC state was found to be $%
\approx 110$ $\mu $eV/unit cell, in agreement with the condensation energy
in the BCS limit\cite{IntJModPhys}.

Figure 1 shows $\sigma _{i}(\omega )$ and $\sigma _{r}(\omega )$ at several
time delays after photoexcitation with a 150 fs pulse with \emph{F} $\sim 3$ 
$\mu $J/cm$^{2}$. Since $\sigma _{i}$ provides a direct probe\cite{Nuss} of
the condensate density, $n_{s}$, by measuring PI changes in $\sigma _{i}$
direct information of $n_{s}$ dynamics can be extracted. On the other hand,
the increase and subsequent recovery of $\sigma _{r}(\omega )$ corresponds
to an initial increase in the number of QPs followed by their recombination%
\cite{THzAveritt}. The decrease of $\sigma _{i}(\omega )$ $-$ corresponding
to a reduction of $n_{s}$ $-$ occurs on the timescale of several ps,
followed by recombination dynamics on the timescale of several hundred ps.
To obtain a more detailed time evolution of the PI changes in $\sigma
(\omega )$, we utilize the fact that the induced changes in the electric
field transient, $\Delta E_{sam}$, are mainly due to a phase shift of the
electric field - see inset to Fig. 1a). The origin of the phase shift is the
so-called kinetic inductance due to SC pairing, i.e. the conductivity due to
the SC pairs is purely imaginary with a $1/\omega $ dependence resulting in
the overall phase shift in $E_{sam}(t)$ below T$_{c}$. When \emph{F} is
smaller than the fluence corresponding to destruction of the SC state, $%
\Delta E_{sam}(t=t_{0})\ $($t_{0}$ is a fixed point of $E_{sam}(t)$ which,
for these experiments, was at the point of maximum time derivative of the
electric field -- indicated by the arrow in inset to Fig.1) is proportional
to the PI conductivity $\Delta \sigma $ (both $\Delta \sigma _{i}$ and $%
\Delta \sigma _{r}$ have the same dynamics - see inset to Fig. 1b).
Therefore by measuring $\Delta E_{sam}(t=t_{0})$ while scanning the pump
line, the PI conductivity dynamics $\Delta \sigma (t)$ can be extracted
rapidly (avoiding the signal drifts associated with the long term laser
stability) and with much higher temporal resolution.

Fig. 2a) presents the induced conductivity dynamics as a function of T (%
\emph{F} $\approx 1$ $\mu $J/cm$^{2}$). The relaxation time $\tau _{R}$,
obtained by fitting the data to $exp(-t/\tau _{R})$, shows a pronounced
T-dependence, plotted by circles in the inset to Fig. 2a). Upon increasing
T, $\tau _{R}$ first decreases, reaches a minimum, followed by a quasi
divergence as T$_{c}$ is approached.

Similar results are obtained from the OPOP experiments - Fig. 2b), where PI
changes in reflectivity ($\Delta R/R$) at optical frequencies are measured
(20 fs pulses at $\hbar \omega \approx 1.54$ eV have been used as a source
of both pump and probe pulses). The normal state dynamics\cite{signchange}
are characterized by a resolution limited risetime (40 fs), followed by a
two-exponential decay, with decay times $\tau _{1}\approx 0.15$ ps and $\tau
_{2}\approx 3.5$ ps (Fig. 2b)). After the initial ps dynamics, $\Delta R/R$
changes sign and the recovery dynamics proceed on a timescale longer than 1
ns. Below T$_{c}$, however, an additional sub-nanosecond response becomes
evident, whose amplitude and recovery dynamics is strongly T-dependent\cite%
{AllOptical}. In particular, the T-dependence of $\tau _{R}$, plotted in the
inset to Fig. 2b), displays the same T-dependence as the OPTP data.

The SC state recovery dynamics do not show any dependence on \emph{F} in the
range $0.1<F<5$ $\mu $J/cm$^{2}$. This suggests that the SC recovery
dynamics are governed by the phonon-bottleneck mechanism, proposed by
Rothwarf and Taylor\cite{RothwarfTaylor} (biparticle recombination should be
intensity dependent). In this case recovery dynamics are governed by the
lifetime of $\omega \gtrsim 2\Delta $ phonons, which are in thermal
equilibrium with QPs.\cite{RothwarfTaylor} Since in MgB$_{2}$ $2\Delta $ is
much smaller than the energy of the lowest optical phonon mode ($\approx $ $%
40$ meV)\cite{Osborn}, the SC recovery is governed by the decay of \emph{%
acoustic} phonons. The decay of $\omega \gtrsim 2\Delta $ phonon population
is governed either by anharmonic decay to $\omega <2\Delta $ phonons or
escape of $\omega \gtrsim 2\Delta $ phonons to the substrate\cite{AllOptical}%
. The fact that no change in $\tau _{R}$ is observed when comparing the data
taken on 80, 100 and 400 nm films suggests that in MgB$_{2}$ the predominant
mechanism that governs the SC condensate recovery is anharmonic phonon
decay. This assignment is further supported by the observed T-dependence of $%
\tau _{R}$, which near T$_{c}$\cite{Kabanov} shows $\tau _{R}\varpropto
1/\Delta (T)$, and the estimated anharmonic decay time\cite{Maris} of the 10
meV longitudinal acoustic phonon in MgB$_{2}$ of $\approx 1$ ns\cite%
{IntJModPhys}.

When discussing the OPOP data we should mention the peculiar ps normal state
dynamics, present also in the SC state. In metals, ps QP dynamics are
usually interpreted in terms of the two-temperature model (TTM) \cite%
{HFPaper,allen,Brorson}. Here the assumption that the e-e scattering is much
faster than the e-ph scattering leads to the description of the PI transient
in terms of the time evolution of the electronic temperature T$_{e}$, i.e. $%
\Delta R(t)=\partial R/\partial T$ $\Delta T_{e}(t)$. The dynamics are due
to e-ph thermalization which is proportional to the e-ph coupling constant $%
\lambda $\cite{HFPaper,Brorson}. For MgB$_{2}$, however, the ps dynamics are
inconsistent with the TTM since: i) the $\Delta R/R$ transient changes sign
with time\cite{signchange}, ii) $\partial R/\partial T$ at 1.5 eV is negative%
\cite{Tu}, so $\Delta R/R<0$ in the TTM, contrary to what is observed, iii)
the expected e-ph thermalization time\cite{Dolgov} is $\tau _{ep}\approx 20$
fs, which is much shorter than the experimental values of either $\tau _{1}$
or $\tau _{2}$, and iv) the initial dynamics do not change upon cooling
below T$_{c}$, where the bottleneck in the relaxation due to the presence of
the SC gap should strongly affect the recovery dynamics\cite%
{Kabanov,RothwarfTaylor}. The above arguments suggest that the origin of the
ps dynamics measured in OPOP experiments is not associated with e-ph
thermalization, and indicates that the assumption of e-e thermalization
being much faster than the e-ph scattering is invalid in MgB$_{2}$. The
coupling between electrons and high frequency optical phonons in MgB$_{2}$
is very strong (in particular the coupling of $\sigma $ band electrons to
the E$_{2g}$ phonon mode at 60-80 meV)\cite{Mazin,Golubov,Choi}. Thus, it is
quite possible that the situation in MgB$_{2}$ is reversed (or at least that
the two timescales are comparable). In this scenario the initial relaxation
of photoexcited electrons proceeds via emission of high frequency (60-80
meV) optical phonons which only subsequently release their energy to the
electron system via phonon-electron scattering, and to low energy phonons
via anharmonic decay. Therefore we argue that the ps dynamics observed in
OPOP experiments is due to the energy relaxation of the optical phonon
population, rather than e-ph thermalization.

Finally, let us discuss the rise-time dynamics, reflecting the Cooper-pair
breaking processes (i.e. the initial reduction of $n_{s}$). Fig. 2a) and the
inset to Fig. 1b) clearly show a finite risetime in the induced change in
conductivity, indicating that it takes some time for the completion of
pair-breaking following optical excitation. Furthermore, Fig. 2a) shows that
the PBD are T-dependent (the PBD becomes faster as T is increased). Also,
the PBD depend on the photoexcitation intensity. Fig. 3a) shows the early
time $\Delta \sigma (t)$ taken at 7K for different fluences. While the solid
symbols represent the data obtained by measuring $\Delta E_{sam}(t)$, the
open symbols represent the induced change in the conductivity obtained
directly using the two-dimensional scanning technique. The agreement between
the two data sets clearly shows that the phase changes accurately reveal the
condensate dynamics. This is particularly important, since measuring $\Delta
E_{sam}(t)$ enables the study of condensate dynamics with sub-picosecond
resolution (limited by optical pulsewidths to $\approx 0.3$ ps), while the
two-dimensional scanning technique is limited by the THz pulse width ($\sim
2 $ ps) \cite{Schmuttenmaer}.

In view of the normal state OPOP data, which suggest that high energy
electrons initially release their energy (or at least a significant portion
of it) via emission of high frequency phonons (which only subsequently break
Cooper pairs), there is a natural explanation of the observed fluence and
T-dependence of the PBD in MgB$_{2}$\cite{AllOptical}. To show this, we
consider the phenomenological Rothwarf-Taylor model\cite{RothwarfTaylor},
where the dynamics of the QP and high frequency ($\omega >2\Delta $) phonon
densities, $n$ and $N$, are described by a set of two coupled differential
equations\cite{RothwarfTaylor}. Since the SC recovery dynamics proceed on a
much longer timescale than the PBD, the term describing the loss of $\omega
>2\Delta $ phonons by processes other than pair excitation can be neglected,
and the equations are given by: 
\begin{equation}
dn/dt=\beta N-Rn^{2}\text{ \ ; }dN/dt=\frac{1}{2}[Rn^{2}-\beta N]  \label{RT}
\end{equation}%
Here $R$ is the bare quasiparticle recombination rate and $\beta $ is the
probability for pair-breaking by phonons\cite{RothwarfTaylor}. With the
initial condition that after photoexcitation (and initial sub-picosecond e-e
and e-ph dynamics) the QP and high frequency phonon densities are $n_{0}$
and $N_{0}$, the subsequent time evolution of $n$ is given by\cite%
{AllOptical} 
\begin{equation}
n\left( t\right) =\frac{\beta }{R}\left[ -\frac{1}{4}-\frac{1}{2\tau }+\frac{%
1}{\tau }\frac{1}{1-K\exp \left( -t\beta /\tau \right) }\right]  \label{Eq1}
\end{equation}%
where $K$ and $\tau $ are dimensionless parameters determined by the initial
conditions: 
\begin{equation}
K=\frac{\frac{\tau }{2}\left( \frac{4Rn_{0}}{\beta }+1\right) -1}{\frac{\tau 
}{2}\left( \frac{4Rn_{0}}{\beta }+1\right) +1}\text{ };\text{ }\frac{1}{\tau 
}=\sqrt{\frac{1}{4}+\frac{2R}{\beta }\left( n_{0}+2N_{0}\right) }.
\label{Eq2}
\end{equation}%
Eq.[\ref{Eq1}] has three distinct regimes, depicted in inset to Fig. 3a). $%
K=0$ corresponds to the stationary solution, when $n_{0}$ and $N_{0}$ after
photoexcitation are already in quasi-equilibrium (at a somewhat elevated T),
i.e. $Rn_{0}^{2}=\beta N_{0}$, and $n(t)$ is a step function. The regime $%
0<K\leq 1$ corresponds to the situation when, following excitation, the
number of QPs is higher than the quasi-equilibrium value, while $-1\leq K<0$
represents the opposite situation when the initial T of $\omega >2\Delta $
phonons is higher than the quasi-equilibrium one. As Fig. 3 shows, the
latter situation seems to be realized in the case of MgB$_{2}$, consistent
with the analysis of the OPOP data.

At low T, when the density of thermally excited QPs and $\omega >2\Delta $
phonons is exponentially small, the change in conductivity is proportional
to the PI quasiparticle density. Therefore we can fit\cite{convolution} the
conductivity dynamics using Eq.(\ref{Eq1}). Best fits to the data taken at
various intensities are plotted by solid lines in Fig. 3a) showing extremely
good agreement with the data. The two fitting parameters $\tau /\beta $ and $%
K$ are plotted in panels b) and c).

Importantly, the microscopic parameters $R$ and $\beta $ can be extracted by
fitting the $\tau /\beta $ and $K$ vs. the absorbed energy density\emph{\ }$%
\Omega $ to Eqs.(\ref{Eq2}). We assume that photoexcitation with $\Omega $
creates $n_{0}=p\Omega /\Delta $ QPs (with energy $\Delta $) and $%
N_{0}=(1-p)\Omega /2\Delta $ high frequency 2$\Delta $ phonons, where $p$ is
the portion of $\Omega $ that initially goes to QPs. A best fit to the two
data sets (dashed lines) is obtained when $p\approx 6\%$ (further supporting
OPOP data analysis) giving $\beta ^{-1}=15\pm 2$ ps, $R=$ $100\pm 30$ ps$%
^{-1}$unit cell$^{-1}$.

In conclusion, we have presented the first femtosecond studies of
photoexcited carrier dynamics in MgB$_{2}$ using both OPTP and OPOP
techniques. The T-dependence of the SC state recovery dynamics have a
similar behavior as for the cuprates\cite{Kabanov,YBCOOver,THzAveritt}. The
main difference is that in the cuprates $\tau _{R}$ is a factor of 100
shorter, which we attribute to their larger gap value, causing the dynamics
(and $\tau _{R}$) to be governed by the lifetime of optical phonons, instead
of acoustic phonons as in MgB$_{2}$\cite{AllOptical}. Furthermore, we have
presented the first observation of time-resolved PBD. The PBD are found to
be temperature and fluence dependent, which is attributed to the initial
creation of high frequency phonons, rather than QPs, and supported by
detailed analysis of the PBD in terms of the Rothwarf-Taylor equations\cite%
{RothwarfTaylor}. Importantly, similar physics may be responsible for the
longer risetimes observed in single layer cuprates ($\approx 500$ fs) \cite%
{Schneider} in comparison to double layered ones ($\approx 0.1$ ps)\cite%
{Kabanov,THzAveritt,YBCOOver}.

Research supported by the US DOE. We wish to thank G.L. Carr, I.I. Mazin, T.
Mertelj, D. Mihailovic, K. A. M\"{u}ller, and J.J. Tu for valuable
discussions.

\subsection{Figure Captions:}

\textbf{Figure 1: }The a) imaginary and b) real conductivity as a function
of frequency shown at various time delays following excitation with a
fluence $\emph{F}\sim 3$ $\mu $J/cm$^{2}$ at 7K. Inset to a): $E_{sam}(t)$
at 7K and 35K. Inset to b): the time evolution of $\sigma _{r}$ and $\sigma
_{i}$ taken at $\nu =0.8$ THz.

\textbf{Figure 2:} a) The PI change in conductivity ($\Delta \sigma \propto
\Delta E_{sam}$) as a function of temperature. b) The PI reflectivity traces
at various temperatures plotted on a semi-log scale. The normal state trace
(90K) is vertically shifted for clarity. The T-dependence of the
superconducting state recovery time, $\tau _{R}$, determined by single
exponential fits to the data (solid lines) is presented in the insets.

\textbf{Figure 3:} a) The rise time dynamics at 7 K taken at various \emph{F}
in $\mu $J/cm$^{2}$ (for presentation purpose all traces have been shifted
vertically by 0.1). Solid circles represent the data obtained by scanning
the pump line while measuring $\Delta E(t=t_{0})$, while the open circles
correspond to $\Delta \sigma $ measured directly. Solid lines are fits to
the data using Eq.(2), with the intensity \emph{F} (absorbed energy $\Omega $%
) dependence of $\tau $ and $K$ plotted in panels b) and c). Dashed lines in
b)\ and c) represent the best fit to Eqs.(3). Inset to panel a): Solutions
of Eq.(2) for different $K$.


\begin{thebibliography}{99}
\bibitem{signchange} The normal state dynamics are virtually T-independent
up to 400 K and did not change when the probe photon energy was tuned from
1.45 eV to 1.65 eV.
\bibitem{AllOptical} A detailed analysis of OPOP data together with
Rothwarf-Taylor analysis will be presented elsewhere.
\bibitem{convolution} To account for experimental resolution of $\sim 0.3$
ps, and the initial e-e and e-ph scattering processes, Eq.[\ref{Eq1}] has
been convoluted with Gaussian of 0.8 ps FWHM.
\end{thebibliography}
\end{document}